# Graphical Abstract

## NRC-Net: Automated noise robust cardio net for detecting valvular cardiac diseases using optimum transformation method with heart sound signals

Samiul Based Shuvo, Syed Samiul Alam, Syeda Umme Ayman, Arbil Chakma, Prabal Datta Barua, U Rajendra Acharya

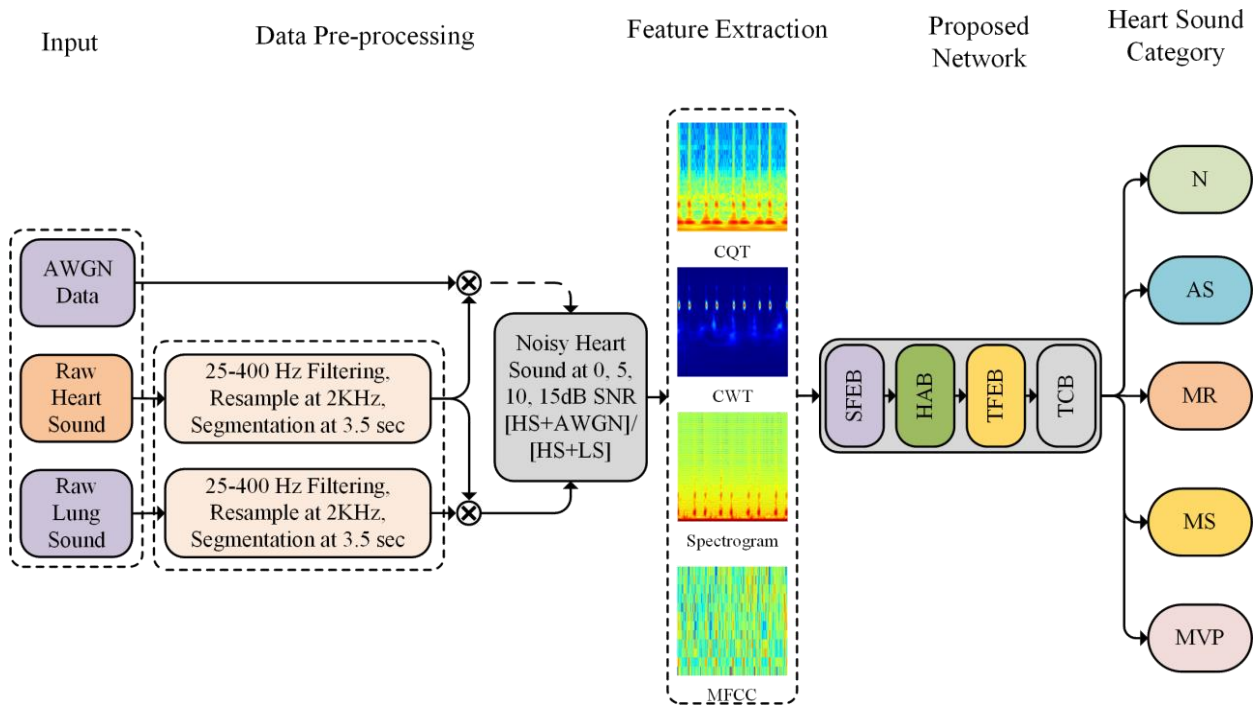

**Graphical Abstract:** A graphical representation of the classification workflow of noisy heart sound (HS). After several pre-processing steps, HS and lung sound (LS) have been acquired framing at 3.5sec and are mixed at specific SNRs to yield noisy HS datasets. Next, in the transformation phase, 2D input images are generated using four transformation techniques (CQT, CWT, STFT and MFCC). Finally, in the classification phase, the 2D images for each category have been successively passed to the proposed architecture. An overview of the NRC-Net architecture consisting of four stages, i.e., Spatial Feature Extractor Block (SFEB), Holistic Attention Block (HAB), Temporal Feature Extractor Block (TFEB), and Terminal Classification Block (TCB) is also demonstrated.

# Highlights

**NRC-Net: Automated noise robust cardio net for detecting valvular cardiac diseases using optimum transformation method with heart sound signals**

Samiul Based Shuvo, Syed Samiul Alam, Syeda Umme Ayman, Arbil Chakma, Prabal Datta Barua, U Rajendra Acharya

- To the best of our knowledge, we are the substantial group to propose a robust deep learning architecture for the classification of noisy heart sounds and obtain the optimal feature transformation method for detecting valvular cardiac abnormalities.

- To enhance the classification performance of the CVDs screening, we introduced a lightweight convolutional recurrent neural network named noise-robust cardio net (NRC-Net) which can detect valvular cardiac disorders in the presence of noisy heart sound signals. The proposed model comprised spatial and temporal feature extraction blocks, extracting both spatial and temporal features from the PCG signal. A holistic attention block has also been integrated to enhance computational efficiency and emphasize more relevant features.

- To extract salient features from the noisy heart sound signal for identifying of various valvular heart diseases, a comprehensive study has been conducted exploiting MFCC, CWT, CQT, and STFT spectrogram using the VGG16 network.

- To develop the robust proposed network, the performance of the model has been evaluated using a 10-fold cross-validation (CV) technique. As a result, the model outperformed with CWT showing the highest classification accuracy of 99.7% on clean data compared to other existing works.

- The proposed model showed superiority over the two well-known state-of-the-art (SOTA) networks i.e. VGG16 and MobileNet V2 architecture in terms of performance and computational overhead.

# NRC-Net: Automated noise robust cardio net for detecting valvular cardiac diseases using optimum transformation method with heart sound signals

Samiul Based Shuvo[a,**], Syed Samiul Alam[b,*], Syeda Umme Ayman[a], Arbil Chakma[b], Prabal Datta Barua[c,d,e,f,g,h,i,j,k,l,m], U Rajendra Acharya[d,m,n]

[a]m-health Lab, Department of Biomedical Engineering, Bangladesh University of Engineering and Technology, Dhaka-1205, Bangladesh
[b]Department of Electronics Engineering, Kookmin University, Seoul 02707, South Korea
[c]Cogninet Australia, Sydney, NSW 2010, Australia
[d]School of Business, University of Southern Queensland, Springfield Central QLD 4300, Australia
[e]Faculty of Engineering and Information Technology, University of Technology Sydney, Sydney, NSW 2007, Australia
[f]Australian International Institute of Higher Education, Sydney, NSW 2000, Australia
[g]School of Science Technology, University of New England, Australia
[h]School of Mathematics, Physics and Computing, University of Southern Queensland, Springfield, Australia
[i]School of Biosciences, Taylor's University, Malaysia
[j]School of Computing, SRM Institute of Science and Technology, India
[k]School of Science and Technology, Kumamoto University, Japan
[l]Sydney School of Education and Social work, University of Sydney, Australia
[m]Department of Biomedical Engineering, School of Science and Technology, SUSS University, Singapore, Singapore
[n]Department of Bioinformatics and Medical Engineering, Asia University, Taichung, 41354, Taiwan
[o]School of Mathematics, Physics and Computing, University of Southern Queensland, Springfield, Australia

**Abstract**

**Objective:** Cardiovascular diseases (CVDs) can be effectively treated when detected early, reducing mortality rates significantly. Traditionally, phonocardiogram (PCG) signals have been utilized for detecting cardiovascular disease due to their cost-effectiveness and simplicity. Nevertheless, various environmental and physiological noises frequently affect the PCG signals, compromising their essential distinctive characteristics. The prevalence of this issue in overcrowded and resource-constrained hospitals can compromise the accuracy of medical diagnoses. Therefore, this study aims to discover the optimal transformation method for detecting CVDs using noisy heart sound signals and propose a noise robust network to improve the CVDs classification performance. **Methods:** For the identification of the optimal transformation method for noisy heart sound data mel frequency cepstral coefficients (MFCCs), short-time Fourier transform (STFT), constant-Q nonstationary Gabor transform (CQT) and continuous wavelet transform (CWT) has been used with VGG16. Furthermore, we propose a novel convolutional recurrent neural network (CRNN) architecture called noise robust cardio net (NRC-Net), which is a lightweight model to classify mitral regurgitation, aortic stenosis, mitral stenosis, mitral valve prolapse, and normal heart sounds using PCG signals contaminated with respiratory and random noises. An attention block is included to extract important temporal and spatial features from the noisy corrupted heart sound. **Results:** The results of this study indicate that,CWT is the optimal transformation method for noisy heart sound signals. When evaluated on the GitHub heart sound dataset, CWT demonstrates an accuracy of 95.69% for VGG16, which is 1.95% better than the second-best CQT transformation technique. Moreover, our proposed NRC-Net with CWT obtained an accuracy of 97.4%, which is 1.71% higher than the VGG16.**Conclusion:** Based on the outcomes illustrated in the paper, the proposed model is robust to noisy data and can be used in polyclinics and hospitals to detect valvular cardiac diseases accurately.

**1. Introduction**

Cardiovascular diseases (CVDs), the predominant factor of global mortality, are killing around 17.9 million human lives annually, representing 2% of global death [1]. Study reveals that developing and underdeveloped regions have high prevalence rate to CVD-related morbidity due to the lack of proper diagnostic equipment, inadequate facilities and insufficient trained medical professionals [2]. Early identification and treatment is essential to reduce the risk factors and unexpected consequences of CVDs like untimely death as well as social burdens. The phonocardiogram (PCG) signal deciphers the mechanical activity of the heart valves, consisting of two fundamental heart sounds, murmurs, and other associated sounds [3] These heart sounds are generated during blood flow into the heart containing important information about the functionality and physiological condition of the cardiovascular system. Thus, an early indication of potential cardiac abnormalities using PCG signals is of paramount significance. The PCG signals are reliable, non-invasive, and comparatively cost effective in the preliminary screening of CVDs. Hence, it is widely used to extract cardiac information from heart sounds and detect abnormalities [4]. In third-world countries, especially in underprivileged regions, the scarcity of trained physicians poses a major challenge as CVDs interpretation from auscultation as it is highly dependent on physicians' expertise, experience and other personal variables. To mitigate these drawbacks, artificial intelligence (AI)-based automatic frameworks are used for the diagnosis and progression management of CVDs [5]. However, objective and reliable assessment of CVDs using computerized methods is quite challenging nowadays due to the interference of physiological and additive noises with heart auscultation. For example, during CVDs screening, the stethoscope captures the breathing sounds along with the heart sound, which causes intrinsic spectral overlap among heart and lung sounds, lowering the CVDs interpretation performance [6]. Moreover, random noises such as hospital ambient noises (phone ringing, door knocking/opening/closing), power line interference, and device variability can affect the PCG's quality and even become the reason of performance degradation of the AI-based automated screening systems in real-life scenarios [7]. Hence, for the accurate assessment of CVDs, the above factors must be addressed carefully while designing a robust auscultation method for medical professionals.

---

*This author contributed equally and share the first authorship.
**Corresponding author
*Email addresses:* sbshuvo@bme.buet.ac.bd (Samiul Based Shuvo), syed.samiul.alam@ieee.org (Syed Samiul Alam), suayman.bme.buet@gmail.com (Syeda Umme Ayman), arbilchakma@ieee.org (Arbil Chakma), prabal.barua@usq.edu.au (Prabal Datta Barua), Rajendra.Acharya@usq.edu.au (U Rajendra Acharya)

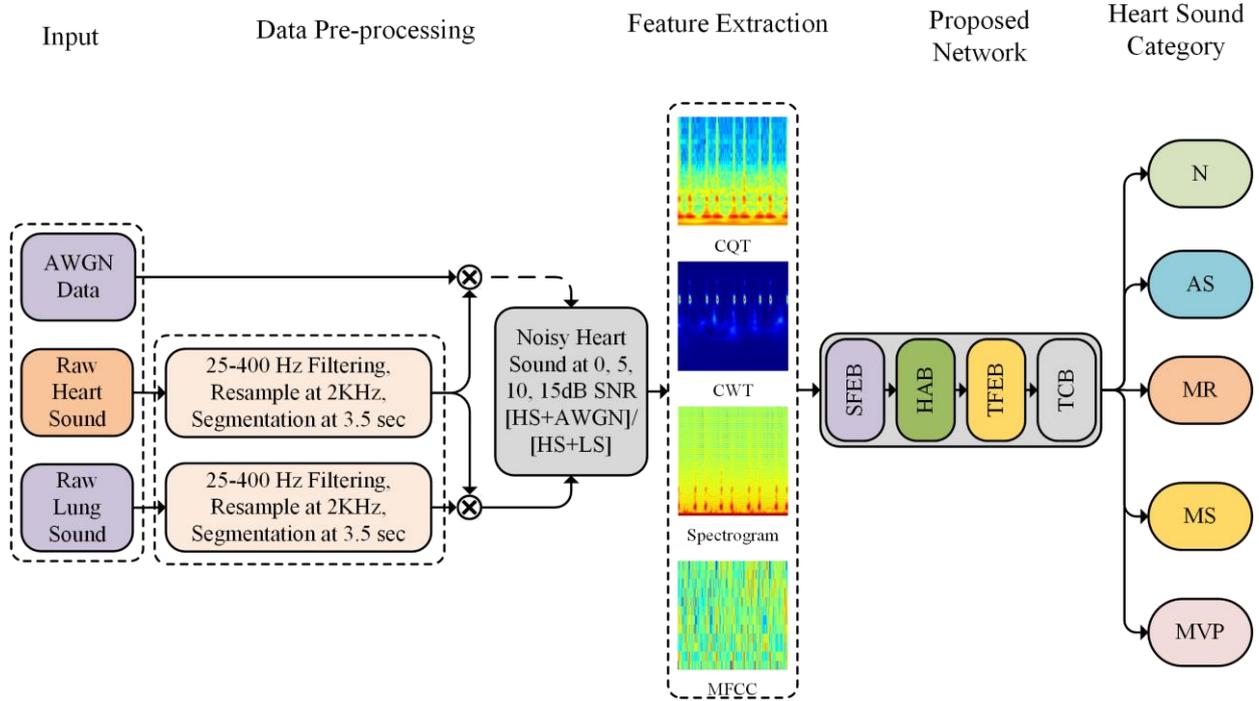

Figure 1: A graphical representation of the classification workflow of noisy heart sound (HS). After several pre-processing steps, HS and lung sound (LS) have been acquired framing at 3.5sec and are mixed at specific SNRs to yield noisy HS datasets. Next, in the transformation phase, 2D input images are generated using four transformation techniques (CQT, CWT, STFT and MFCC). Finally, in the classification phase, the 2D images for each category have been successively passed to the proposed architecture. An overview of the NRC-Net architecture consisting of four stages, i.e., Spatial Feature Extractor Block (SFEB), Holistic Attention Block (HAB), Temporal Feature Extractor Block (TFEB), and Terminal Classification Block (TCB) is also demonstrated.

In the era of AI, a considerable amount of work has been done on automatic heart sound analysis, and CVD detection has been performed on open-access PCG datasets [8, 9, 10]. Several initiatives have been taken for heart sound abnormality detection using these datasets. In most of the cases, time-frequency, and statistical features [11], mel-frequency cepstral coefficients (MFCC) [12, 13, 14, 15] and continuous wavelet transform (CWT) [16, 17] based features have been used. MFCC and mel-spectrogram are the most common among all these. Machine learning (ML)-based classifiers namely k nearest neighbor (kNN) [13], random forest (RF) [11, 12, 13], support vector machine (SVM) [13, 18, 19], and multilayer perceptron (MLP) [13, 17, 18, 20] have been used for the detection of CVDs. But ML-based algorithms depend on hand-crafted features, generating biases in the classification task. On the other hand, deep learning (DL)-based methods provide both generalization and high accuracy. DL-based approaches like recurrent neu-ral networks (RNN) [21] and 1D and 2D CNNs [15, 19, 22, 23, 24, 25, 26] have been widely utilized for the automated detection of CVDs.

Furthermore, most of the work emphasizes on binary classification (normal vs. abnormal) and does not consider the presence of convolutional or additive noises except [19] that generated synthetic data considering some inessential noises using PhysioNet dataset[10] which itself consist of several noises (e.g., breathing, stethoscope movement, intestinal activity, peripheral talking, etc.) in the recordings and thereby looses the reliability of the classification task . Hence, there is a lack of practical considerations and reliability in the automatic detection of CVDs. Besides, the classification of CVDs with noisy data goes through the denoising process using conventional techniques making the whole classification task more challenging in the real-life scenario as these denoising procedures are computationally intensive and require additional memory.

The main contributions of this paper are:

- To the best of our knowledge, we are the substantial group to propose a robust deep learning architecture for the classification of noisy heart sounds and obtain the optimal feature transformation method for detecting valvular cardiac abnormalities.

- To enhance the classification performance of the CVDs screening, we introduced a lightweight convolutional recurrent neural network named noise-robust cardio net (NRC-Net) which can detect valvular cardiac disorders in the presence of noisy heart sound signals. The proposed model comprised spatial and temporal feature extraction blocks, extracting both spatial and temporal features from the PCG signal. A holistic attention block has also been integrated to enhance computational efficiency and emphasize more relevant features.

- To extract salient features from the noisy heart sound signal for identifying of various valvular heart diseases, a comprehensive study has been conducted exploiting MFCC, CWT, CQT, and STFT spectrogram using the VGG16 network.

- To develop the robust proposed network, the performance of the model has been evaluated using a 10-fold cross-validation (CV) technique. As a result, the model outperformed with CWT showing the highest classification accuracy of 99.7% on clean data compared to other existing works.

- The proposed model showed superiority over the two well-known state-of-the-art (SOTA) networks i.e. VGG16 and MobileNet V2 architecture in terms of performance and computational overhead.

The proposed framework for NRC-Net is demonstrated in Figure 1. The rest of this article is organized as follows. In section 2 we have elaborated the literature review. Section 3 overviews the datasets, pre-processing steps and transformation methods applied in this work. Section 4 provides a detailed description of the proposed architecture. Section 5 discusses the evaluation criteria and experimental results with a comparison with the existing literature. Section 6 explains the future directions of the work along with the superiority and drawbacks. Finally, we conclude our study in section 7.

Table 1: Summary of works carried out on automated detection of heart sounds using GitHub PCG dataset [9].

| Author | Transformation Method | Classifier | Accuracy% |
|---|---|---|---|
| Yaseen et al. [9] (2018) | Discrete wavelet transform (DWT) and MFCC | 1. SVM<br>2. DNN<br>3. kNN | 1. **97.90**<br>2. 92.10<br>3. 97.40 |
| M. Alqudah et al. [27] (2019) | Instantaneous frequency (Statistical features) | 1. RF<br>2. kNN | 1. 95.00<br>2. 95.00 |
| M. Alqudah et al. [28] (2020) | 1. Full bi-spectrum<br>2. Contour bi-spectrum | AOCTNet | 1.**98.70**<br>2. 97.10 |
| Ghosh et al. [29] (2020) | Spline kernel-based Chirplet transform<br>1. L1-norm<br>2. Sample entropy<br>3. Permutation entropy | Deep layer kernel sparse representation network (DLKSRN)<br>1. Holdout<br>2. 10-fold CV | 1. 99.23<br>2. **99.24** |
| Oh et al. [30] 2020 | - | WaveNet | 94.00 |
| Baghel et al. [31] (2020) | - | 1D CNN<br>1. Augmented data<br>2. Non-augmented data | 1. **98.60**<br>2.96.23 |
| Zeng et al. [32] (2021) | Tunable Q-factor wavelet transform and fast and adaptive multivariate empirical mode decomposition<br>1. Shannon energy envelop (SEE) | Supervised RBF neural networks | 98.48 |
| Shuvo et al. [33] (2021) | - | CardioXNet | 99.60 |
| M. Alkhodari et al. [34] (2021) | 1D wavelet smoothing | CNN-BiLSTM | 99.32 |
| Tiwari et al. [35] (2021) | 1. Constant-Q Transform (CQT)<br>2. Variable-Q Transform(VQT)<br>3. Hybrid Constant-Q Transform(HCQT)<br>4. MFCC | ConvNet | 1. 94.00<br>2. 94.00<br>3. 93.00<br>4. **96.00** |
| Kobat et al. [36] (2021) | Improved 1D binary pattern (IBP) | 1.KNN<br>2.SVM | 1. **99.50**<br>2. 98.30 |
| Y. Al-Issa et al. [37] (2022) | - | Hybrid light CNN-LSTM<br>1. Augmented Data<br>2. Non-augmented Data | 1. **99.87**<br>2.98.48 |

## 2. Related Works

As mentioned earlier, many researchers sought to discriminate various CVDs using PCG signals employing ML and DL-based frameworks using publicly available datasets and private datasets. In this section, we summarized an overview of research works in this domain using the Github PCG dataset [9] (presented in Table 1).

It may be noted from the table that, all these works have been conducted on the noise-free clean PCG data. The performance of such models will alter in noisy actual real-world scenarios.

To address this issue, our proposed NRC-Net presents a noise-robust DL-based solution for the classification and detection of CVDs. This approach is crucial for developing effective and reliable remote healthcare devices that can accurately detect CVDs, especially in areas with limited access to healthcare resources is limited.

## 3. Materials and methods

In this section, an overview of the datasets, signal pre-processing stages, and different transformation methods applied in this work are provided.

### 3.1. Dataset(s)

The datasets used for this work is explained in this subsection.

#### 3.1.1. Heart sound dataset

The PCG dataset used in this work comprises 5 classes: mitral regurgitation (MR), aortic stenosis (AS), mitral stenosis (MS), mitral valve prolapse (MVP), and normal (N) [9]. There are 1000 audio files available in the database, with each signal in this dataset consisting of 3 complete cardiac cycles of noise-free heart sound sampled at 8KHz.

#### 3.1.2. Lung sound dataset

The International Conference on Biomedical Health Informatics (ICBHI) 2017, a benchmark dataset of lung auscultation sounds that is publicly available was used in this work. The dataset was accumulated by 2 research groups from Greece and Portugal, containing a total recording of 5.5 hours, sampled at 4KHz, 14KHz, and 44.1KHz with annotated respiratory cycles from 126 patients with 920 recordings [38].

### 3.2. Data prepossessing

This subsection presents the major steps involved in data preprocessing.

#### 3.2.1. PCG signal preparation

The PCG signals have been filtered with a bandpass filter at 50Hz-800Hz and resampled at 2000Hz [39]. The variation in the length of the heart sound files restrains them from being used in any classification algorithm. Data framing is used to solve this irregular length issue. Each signal is converted into a 3.5sec signal with the padding method proposed in [40].

#### 3.2.2. Lung sound preparation

The frequency range of the lung sound signals is 50 Hz-2500 Hz [41]. Hence, the lung sounds are filtered by leveraging Butterworth bandpass filter of order 6. Subsequently, all the signals are resampled to 2000Hz to ensure consistency and normalized to the range [-1,1].

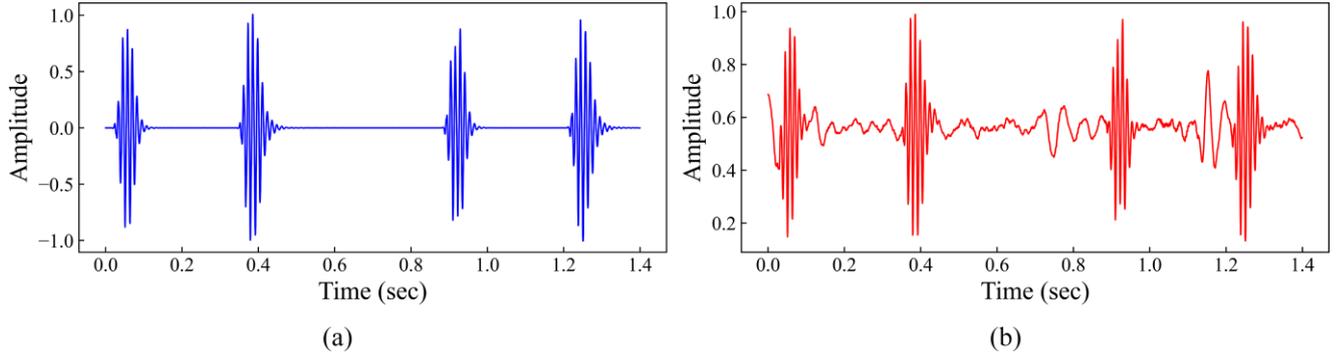

Figure 2: Waveform of (a) clean heart sound (normal), and (b) the same sound corrupted by lung sounds.

*3.2.3. Noisy heart sound signal preparation*

A noisy PCG dataset is used to investigate the noise-robustness of our proposed network. The heart sound samples in this dataset are corrupted by incorporating both lung sound and AWGN to form two noisy datasets (see Figure 2). In each PCG sample, the lung sound and AWGN noise are added at signal-to-noise ratios (SNRs) at different levels (0dB, 5dB, 10dB, and 15dB).

*3.3. Transformation methods*

In this study, we have used CWT, CQT, STFT, and MFCCs to transform the one-dimensional PCG signal into the images with a resolution of 224 x 224 pixels.

*3.3.1. Continuous wavelet transform (CWT)*

CWT is a mathematical method that decomposes a signal with finite energy, x(t), into an orthonormal basis of wavelets, which consists of the mother wavelet, g(t) as well as the dilated and scaled versions of it [42]. The mathematical expression of CWT can be formulated as shown in Eqn. (1):

$$Z(a, b) = \int x(t) * g(t)(\frac{t-a}{b}) \tag{1}$$

where a denotes the scale factor and b is the time location. Low-frequency and high-frequency information are demonstrated by larger and smaller scale values, respectively [43]. The scalogram is the squared modulus of the CWT's coefficient Z [44]. For this work, we have utilized Morse analytic wavelet as the mother wavelet to decompose the heart sound samples into the wavelet domain, with the symmetry parameter and time-bandwidth product of 3 and 60, respectively [45]. The maximum and minimum scales are automatically ascertained by exploiting 10 voices per octave, depending on the wavelet's energy range in frequency and time.

*3.3.2. Mel frequency cepstral coeflcients (MFCCs)*

MFCC is a highly efficient method for extracting a wide range of features obtained through a cosine transformation process. These consine transformations of the real logarithmic short-term energy spectrum demonstrated on a mel-frequency scale into a set of cepstrum coefficients [46]. The relation between the frequency, f, and the mel-frequency, m is defined by Eqn (2):

$$m = 2595 \log(1 + \frac{f}{700}) \qquad (2)$$

In this work, with a 30 ms hamming window and a step size of 10 ms, we have run overlapping sliding windows over all the heart sound samples to extract the coefficient. Then, we stacked all the co-efficient to generate MFCC plot of each heart sound sample.

### 3.3.3. Short-time Fourier transform (STFT)

An STFT algorithm conserves information in the spectro-temporal domains. It is one of the most widely used and reliable methods for extracting features from a 1D acoustic sample [47]. The STFT is calculated by Eqn (3)

$$STFT(\tau, w) = \int y_a(t) W_a(t - \tau) e^{(-jwt)} dt \qquad (3)$$

where $W_a(t)$ is denoted as window function and $y_a(t)$ is the signal to be transformed. We used STFT with a Han window length of 128 and a hop length of 64 on all of the heart sound samples to generate an STFT spectrogram. It (spectrogram) is a demonstration of a 1D signal in the time and frequency domain that represents high-frequency and low-frequency components [48].

### 3.3.4. Constant-Q nonstationary Gabor transform (CQT)

A spectro-temporal representation known as the CQT, has geometrically spaced frequency bins and uniform Q-factors (ratios of center frequencies to bandwidths) across all bins. Since the CQT is essentially a wavelet transform, the frequency resolution is higher for low frequencies and the temporal resolution is higher for high frequencies [49]. We set the number of bins per octave to 12 with a Hann window size of 128 and hop length of 64 to generate a logarithmic STFT spectrogram scale for each heart sound samples.

## 4. Proposed lightweight CRNN architecture

The proposed NRC-Net architecture has been developed, by integrating three novel feature extraction stages for extracting spatial and temporal salient features from the generated input image to classify PCG signals effectively. The first stage consists of a regular deep CNN-based spatial initial feature extractor block (SFEB) for extracting abstract feature representations from the input image. The second stage contains the holistic attention block (HAB) for emphasizing the conspicuous features through channel, spatial and pixel-wise re-calibration and generalization. Finally, in the third stage, the temporal feature extractor block (TFEB) has a single long short-term memory (LSTM) layer to acquire the temporal representations. All the extracted features are then fed to the terminal classification block (TCB) consisting of fully connected layers which are followed by the softmax layer to attain the final class prediction. The detailed discussion of different architectural blocks and sub-modules of the proposed NRC-Net (The high-level network architecture is portrayed in Figure 1) is given below.

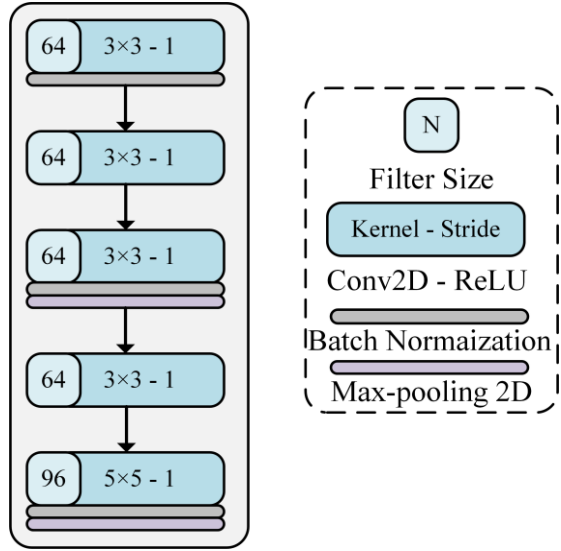

Figure 3: Architecture of spatial feature extractor block (SFEB).

*4.1. Spatial feature extractor block (SFEB)*

The first stage of the NRC-Net is implemented through the SFEB with six consecutive convolutional layers. First, the 3-channel 224 x 224 images are fed into the input layer and passed to the first convolutional layer for further processing. Then, using varying filters and kernel sizes, the convolutional layers convolved the input with 2D kernels for generating abstract feature maps.

A batch normalization layer follows a convolutional layer for stabilizing and speeding up the training process and a max-pooling layer for attaining a translation-invariance effect with reduced network parameters and overfitting. In all the convolutional layers, rectified linear unit (ReLU) is used to introduce non-linearity and attenuate the vanishing gradient issue while ensuring faster convergence by avoiding neuron saturation [50], which is defined as f(y) = max(0, y). Next, the output of this SFEB is fed into the HAB for learning more robust information from the intermediate feature representations. Finally, With the hyperparameter tuning, the hidden state is chosen. (see Figure 3).

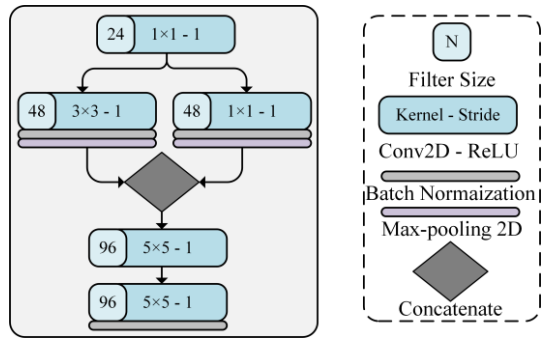

Figure 4: Architecture of holistic attention block (HAB).

*4.2. Holistic attention block (HAB)*

The fundamental component of HAB comprises a squeeze layer with a 1x1 convolution layer and an expanded layer that incorporates parallel convolution layers of sizes 1x1 and 3x3, followed by two successive convolutional layers [51]. In the subsequent expanding layer, a concatenation operation merges these two convolution layers. Once the squeeze layer produces the output, it goes through an excitation layer that trains itself to emphasize important features. As a result, the squeeze layer diminishes the depth of feature maps, thereby enhancing the proposed models' computational efficiency, and making it suitable for environments with limited resources, including mobile and embedded devices (see Figure 4).

*4.3. Temporal feature extractor block (TFEB)*

The TFEB block is formed using two parallel LSTM layers concatenated together. The main motivation for constructing this block is to simultaneously utilize multiple parallel LSTM units to process diverse temporal cues of varying resolutions in the reshaped extracted features of the HAB [52]. LSTM units with recurrent hidden states can handle sequential inputs and retain temporal features that get eliminated in feedforward neural network structures like CNN, which consider all inputs independent of one another. In NRC-Net, we have used two consecutive TFEBs with 64 and 32 hidden layers. With the help of hyperparameter tuning, the hidden state is chosen (see Figure 5).

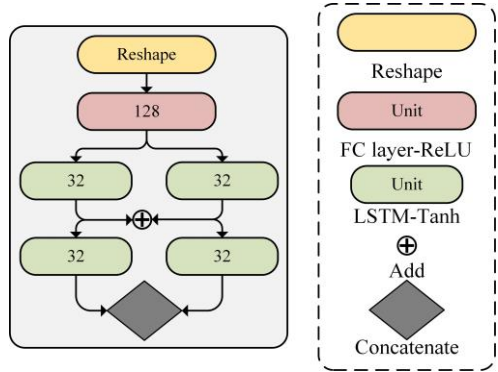

Figure 5: Architecture of temporal feature extractor block (TFEB).

*4.4. Terminal classification block (TCB)*

The TCB is formed using the fully connected and softmax layers fed by the TFEB. First, the extracted feature vector is flattened and fed into 5 fully connected layers, afterward with an output layer with the probability nodes for each class. Next, the softmax function calculates each probability value, which generates a vector with values in the range [0, 1] and denotes a categorical probability distribution over the five classes. To minimize the over-fitting, dropout regularization has been employed following each fully connected layer (see Figure 6).

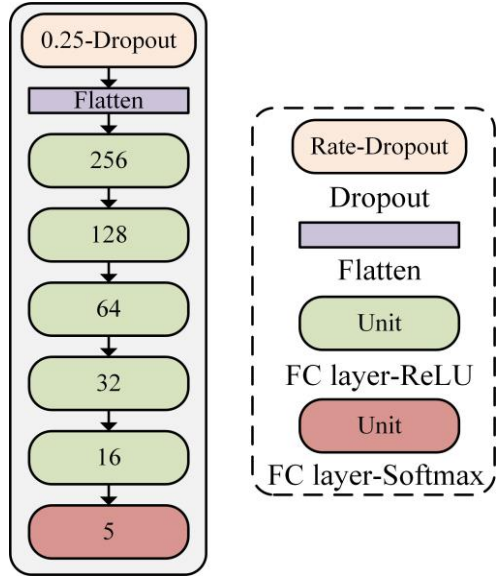

Figure 6: Architecture of terminal classification block (TCB).

## 5. Experimental results

The evaluation criteria used and results obtained using our proposed model are presented in the following subsections.

*5.1. Evaluation criteria*

The accuracy, sensitivity, and specificity evaluation matrices are used to assess the model's performance. [53] The ratio of the correct predictions of model to the total predictions is defined as accuracy and it can be calculated by Eqn. (4).

$$Accuracy = \frac{T_P + T_N}{T_P + F_P + T_N + F_N} \quad (4)$$

Sensitivity, also known as recall is the ratio of true observations that are classified correctly to all the observations in that class. It can be calculated by Eqn. (5).

$$Sensitivity = \frac{T_P}{T_P + F_N} \quad (5)$$

Specificity is defined as negative observations made by a model that is correct to all other negative observations and can be evaluated by Eqn. (6).

$$Specificity = \frac{T_N}{T_N + F_P} \quad (6)$$

where true positive, true negative, false positive, and false negative are denoted as $T_P$, $T_N$, $F_P$, and $F_N$ respectively.

Table 2: Details of hyperparameters used for the models.

| Hyper-parameters | Values |
|---|---|
| Training data shape | 4000, 224, 224, 3 |
| Test data shape | 5 ∗ (200, 224, 224, 3) |
| Batch size | 16 |
| Learning rate | 0.0001 |
| Epoch | 60 |
| Optimizer | Adam |
| Loss function | Categorical-cross entropy |

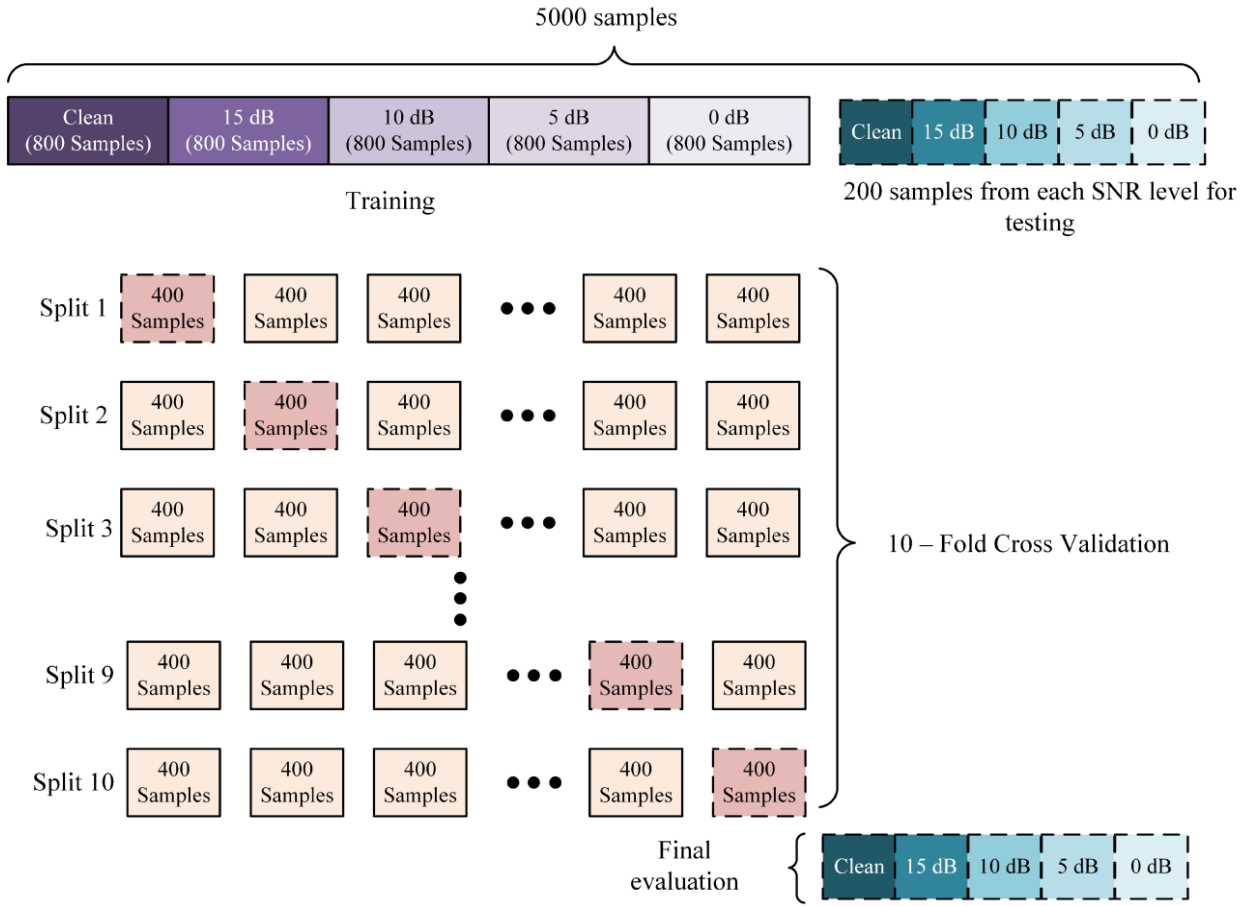

Figure 7: A schematic representation of K-fold CV when K = 10. The initial training dataset contains 4000 samples and is randomly divided into K separate sets. In each iteration of the process, K-1 of these sets is used to train a model (highlighted in light orange), while the remaining set is employed for validation (highlighted in pink). This process is repeated K times, covering all possible combinations of validation sets. The final evaluation of the model is conducted on 1000 samples at each SNR level, as represented by the blue gradient.

The models utilized in this work are created using Keras and TensorFlow 2.0 and trained using an Intel Xeon 3.8GHz CPU, 256 GB RAM, and an Nvidia Quadro 6000 GPU with 24 GB VRAM. 80% and 20% of the total samples for each SNR level were respectively allocated to the training and testing sets. Furthermore, the model was validated using a 10-fold CV approach, with 10% of the training data held out for validation in each fold(detailed representations are shown in 7). Once the validation process was completed, the best-performing model was chosen and applied to the testing set for inference. The details of hyperparameters used for the proposed model are shown in Table 2.

*5.2. Results*

*5.2.1. Performance on the baseline model*

In this study, the baseline network, i.e., the VGG16 model's performance, is investigated for different spectral transformation techniques. Table 3 illustrates the classification accuracy of VGG16 model for CWT, CQT, MFCC, and STFT spectrograms using different SNRs with lung sound as noise. From this table, it can be seen that CQT and CWT performed better than MFCC and STFT spectrograms. The performance of the baseline model is comparable with CQT and CWT when the SNR is higher. The accuracy of the model for CWT with clean data is 99.5%. The average accuracy for CWT with noisy heart sound is 95.69%. There is a decrease in accuracy as the SNR decreases. On the other hand, despite achieving 99.14% accuracy for CQT with clean data, its performance rapidly declines as the heart sound signal becomes noisier. For example, with CQT, the average accuracy of all noise levels for the baseline model is 93.69% which is 1.95% lower than CWT. Similarly, the average accuracy obtained with the noisy signal's STFT spectrogram and MFCC is 2% and 10% lower than CWT, respectively. The confusion matrix obtained for the VGG16 model with CWT and clean data is shown in Figure 8(b).

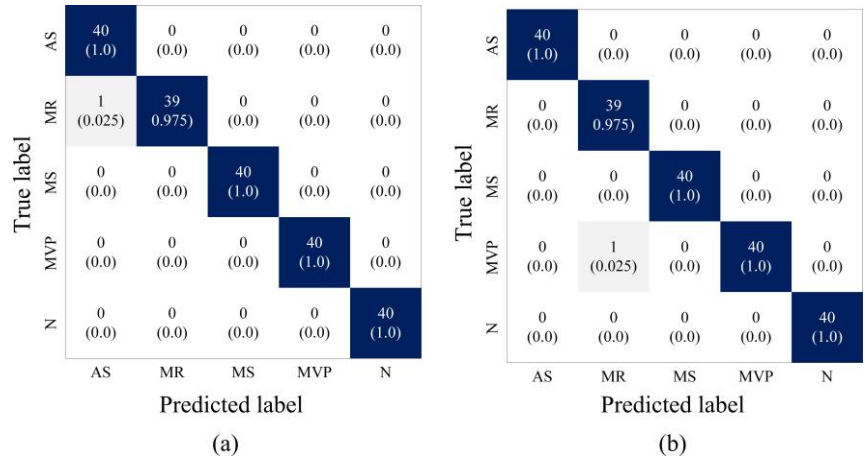

Figure 8: Confusion matrices for the (a) proposed model with CWT(b) VGG16.

*5.3. Classification performance of the proposed framework*

Table 3: Comparison of accuracy(%) for VGG16 model with different transformation techniques.

| SNR | CWT | CQT | Spectogram | MFCC |
|---|---|---|---|---|
| **0 dB** | **92.50** | 87.52 | 82.44 | 73.80 |
| **5 dB** | **95.02** | 91.40 | 92.40 | 78.00 |
| **10 dB** | **95.42** | 94.57 | 97.43 | 89.00 |
| **15 dB** | **96.01** | 96.12 | 97.89 | 92.50 |
| **Clean data** | **99.50** | 99.14 | 98.28 | 95.00 |

To improve the performance of heart sound classification, we have proposed NRC-Net architecture in this work. Table 4 displays the results of 10-fold CV for this proposed architecture. During this process, the entire dataset was divided into ten folds, with one-fold used for validation. The remaining folds used for training in each iteration (see Figure 7). The proposed network achieved almost perfect validation accuracy on the GitHub dataset through 10-fold CV. Furthermore, our model has yielded high sensitivity and specificity scores to support its generalization ability. Additionally, Figure 10(a) illustrated the rapid convergence of the model, with training accuracy reaching 100% and validation accuracy reaching 99.7% within 60 epochs.

Further comparison was performed between our proposed network and the two SOTA architectures, VGG16 and MobileNet V2, in terms of parameters.

Table 4: Ten-fold CV scores obtained for the proposed model.

| Fold | Accuracy(%) | Sensitivity(%) | Specificity(%) |
|---|---|---|---|
| 1 | 99.75 | 100 | 100 |
| 2 | 100 | 100 | 100 |
| 3 | 100 | 99.9 | 99.87 |
| 4 | 100 | 100 | 100 |
| 5 | 100 | 99.79 | 99.86 |
| 6 | 99.75 | 99.73 | 99.88 |
| 7 | 100 | 100 | 100 |
| 8 | 100 | 99.85 | 99.82 |
| 9 | 99.75 | 99.78 | 99.98 |
| 10 | 100 | 100 | 100 |
| **Avg.** | **99.924** | **99.923** | **99.941** |

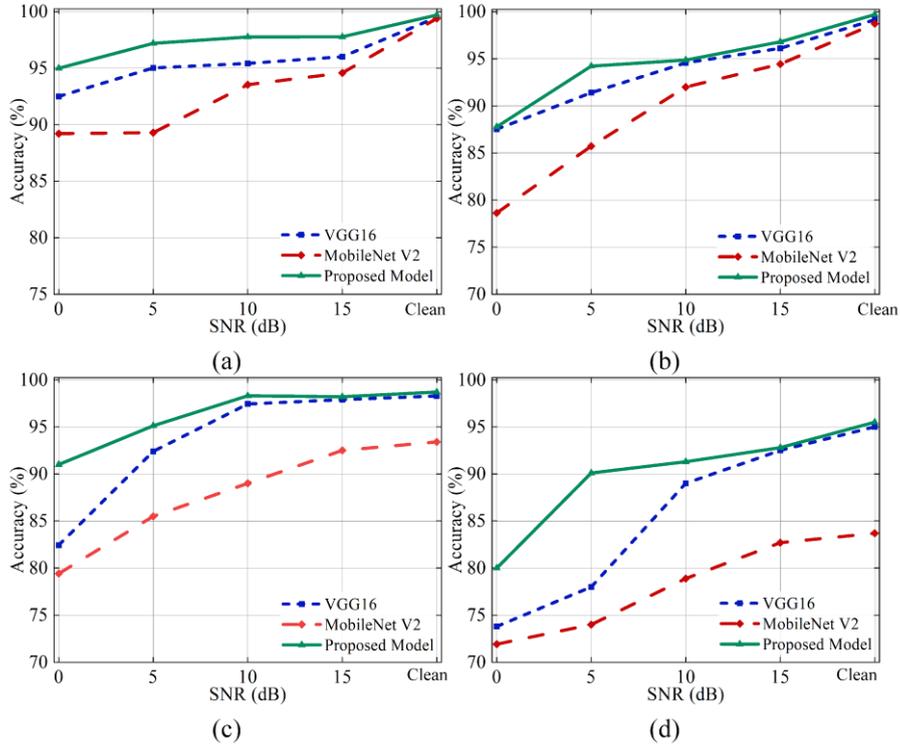

Figure 9: Performance comparison with different transformation methods: (a) CWT, (b) CQT, (c) Spectogram, and (d) MFCC.

Figure 9 illustrates the accuracy of the models on different noise levels. The MobileNet V2 obtained the lowest accuracy with different noise levels. The accuracy of MobileNet V2 for CWT is 99.4% on the clean test data, while both VGG16 and the proposed model showed remarkable results, exceeding 99.5% with CWT. This indicates that both models can achieve a high level of accuracy when evaluated using clean test data. However, our proposed model exhibited slight improved in performance with the VGG16 architecture in this case (see Figure 9). However, when the level of noise is increased, the difference in performance between the models becomes more apparent. The proposed model demonstrated superior performance in all scenarios, compared to the VGG16 and MobileNet V2 models. This superiority can be attributed to the improved capability of the proposed model in extracting relevant features from the noisy data. In the proposed model, we used a squeeze-excitation block that can adaptively reconfigure the weights and suppress irrelevant features, thus providing attention. Hence significantly assisting the learning of important features form the noisy heart sounds. Moreover, the proposed model integrated with the parallel LSTM layers can capture pseudo-periodic temporal features of the heart signals which assist the classification of different valvular diseases which is crucial in identifying the normal class. Figure 8(a) depicts the confusion matrix of the proposed CWT model with clean data.

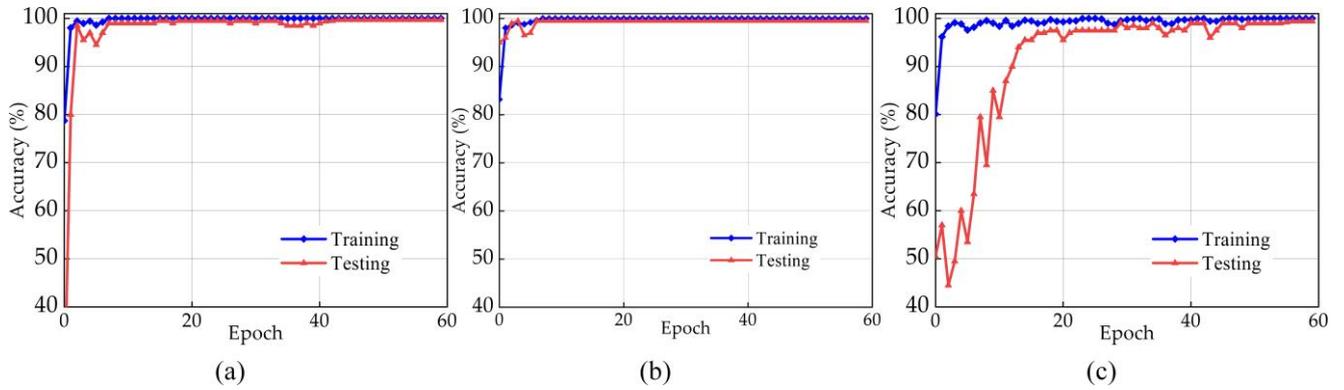

Figure 10: Accuracy vs. epoch curves obtained for: (a) proposed network, (b) VGG16, and (c) MobileNet V2 during training and testing.

## 5.4. Performance on the AWGN

Additive white Gaussian noise (AWGN) is a widely employed type of noise in various research studies to exhibit noise robustness of deep learning models [54, 55].Consequently, to demonstrate our proposed model's superiority and noise robustness, an experiment is conducted using a heart sound dataset corrupted with AWGN. Figure 11 shows the performance of the models on the CWT data for different noise levels. As expected, the proposed model outperformed the other two models for all noise levels. When the noise level is set to 0 dB, the accuracy of the proposed network was 51%, surpassing the accuracy of 47% achieved by the VGG16 model. This trend is observed across all noise levels, proving that the proposed model outperformed the baseline model.

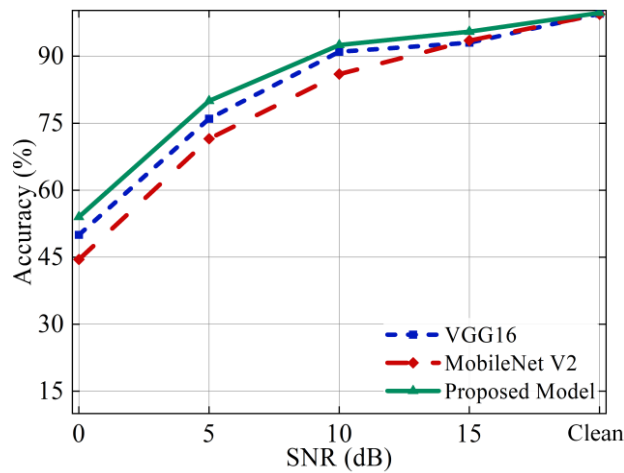

Figure 11: Performance comparison of different models with CWT on AWGN.

## 6. Discussion

### 6.1. Comparison with state-of-the-art (SOTA) techniques

In this study, we have identified heart conditions using the proposed NRC-Net and other established deep learning models utilizing different transformation techniques in the presence of noise. The models were trained and evaluated on the GitHub dataset. Table 5 compared the suggested model's performance metrics with the SOTA methods. All the articles mentioned in Table 5 utilized the GitHub PCG database. It is clear from the table that NRC-Net, despite training on a noisy dataset, obtained promising performance compared to the existing works. None of the previous studies were conducted on a noisy heart sound signal. Therefore, only the results obtained from the clean noise free data have been compared. Authors in [9] used a SVM and RF-based classifiers [27] and reported the accuracies of 97.90% and 94.80% respectively. With its superior feature extraction capability, the proposed model has acquired an overall accuracy of 97.4%, which is 2.6% higher than both of these works. On the contrary, the accuracy of the proposed network on the clean noise free data is 99.70% which is 1.22% and 0.10% higher than [37] and [33], respectively. Our model obtained a sensitivity and specificity of 99.58% and 99.66%, which is 1.6% and 1.2% higher than [37], respectively.

Table 5: Comparison of NRC-Net with existing works using GitHub PCG dataset [9].

| Classifier | Transformation Technique | Noise consideration | Accuracy (%) | Sensitivity (%) | Specificity (%) |
|---|---|---|---|---|---|
| SVM [9] | MFCC + DWT | No | 97.90 | 98.20 | 99.40 |
| RF [27] | Spectogram | No | 94.80 | 94.78 | - |
| DLKSRN [29] | Chirplet Transform | No | 99.24 | - | - |
| Composite classifier [28] | Bispectrum | No | 98.33 | 98.33 | - |
| CNN [31] | - | No | 98.60 | - | - |
| WaveNet [30] | - | No | 94.00 | 92.50 | 98.10 |
| CNN-BiLSTM [34] | - | No | 99.32 | 98.30 | 99.58 |
| CardioXNet [33] | - | No | 99.60 | 99.52 | |
| CNN + LSTM [37] | - | No | 98.48 | 98.52 | 99.58 |
| **NRC-Net** | **CWT** | **No** | **99.70** | **99.58** | **99.60** |
| (Proposed network) | **CWT** | **Yes** | **97.40** | **98.10** | **98.90** |

### 6.2. Comparison of computational performance of the network

Due to limitations in processing and storage space for weights of the filters and parameters, embedded devices are unsuitable for deploying certain models [56]. As a result, computationally intensive training is typically outsourced to cloud computing platforms, which require higher RAM [57]. To address these challenges, lightweight CNN models have become popular among researchers due to their compact size, faster performance, and comparable accuracy to larger deep learning networks. Table 6 depicts the number of parameters for all three models. The proposed model requires fewer parameters than the VGG16 and MobileNet V2 architecture while persisting in better accuracy performance. Moreover, the proposed model's inference time is lower, making it a lightweight network and computationally less expensive as it extracts features more efficiently.

Table 6: Number of parameters used in different models.

| Architecture | Number of Parameters | Inference Time (sec) |
|---|---|---|
| VGG16 | 14,840,133 | 0.003049 |
| MobileNet V2 | 2,571,589 | 0.002738 |
| **NRC-Net** | **1,958,909** | **0.002693** |

The major advantages of the proposed system are as follows:

- Several transformation techniques were evaluated to identify the best transformation technique for cardiac abnormality identification.

- Proposed NRC-Net used a combination of temporal and spatial feature extractors enabling the system to learn from both types of features. This approach is developed to amplify the overall learning and performance of the system by utilizing relevant features.

- Generated model incorporates a squeeze and excitation architecture that allows a more streamlined parameter count than other networks such as VGG16 or MobileNet V2. Using this architecture, the model can be designed to be lightweight and more efficiently deployable on edge devices. This approach is expected to enhance the model's overall practicality and accessibility.

- The most comprehensive study of the PCG signal under the influence of noise and obtaining high classification

- Despite utilizing fewer parameters than traditional models the proposed architecture achieved superior performance compared to the SOTA methods even when operating in noisy environment.

The disadvantages of the proposed system are as follows:

- Employed a two-dimensional feature representation that introduces additional complexity.

- Influence of hospital ambient noise was not considered.

- Due to the lack of a noisy HS dataset we could not validate our proposed system on inherently noisy signals.

In this article, we evaluated the proposed model's performance in classifying heart sound corrupted with lung sound noise and AWGN. The generated model is designed to provide the highest accuracy despite a lightweight architecture that reduces computational complexity and hence lower latency. In the future, we aim to investigate the network's performance in the presence of hospital ambient noise and reconfigure the NRC-Net to accomplish the same task by exploiting raw 1D heart sound data. This will allow us to integrate our proposed NRC-Net with digital stethoscopes or handheld devices powered by cloud server connection in the context of 5G e-health to perform automatic heart condition classification and predict different CVDs accurately. Utilizing our pre-trained network in real-time has the potential to aid medical practitioners in making diagnostic decisions, ultimately leading to substantial benefits for remote e-health services powered by 5G technology. Furthermore, we aim to evaluate the explainability of our proposed model to develop a transparent and interpretable end-to-end system for cardiac anomaly detection without any potential biases. For this purpose, we intend to use gradient-weighted class activation mapping (GradCAM) because the GradCAM method strives to maintain the CNN model's interpretability while preserving the complexity [58].

## 7. Conclusion

In this work, we introduced a lightweight DL-based framework to detect heart conditions at an early stage to address the concern related to premature mortality due to various valvular diseases. We evaluated different transformation techniques for identifying heart conditions in noisy heart sound conditions using the VGG16 model. We found that CWT performed the best compared to other transformation techniques, with an overall accuracy of 95.69% uisng VGG16 architecture. This is nearly a 2% improvement more than the second-best CQT transformation technique. Furthermore, our proposed NRC-Net model improved the classification performance of CWT and other transformation methods and outperformed VGG16 and MobileNet V2. The proposed model demonstrated a 1.84% and 4.33% improvement in overall accuracy with CWT over VGG16 and MobileNet V2, respectively. The evaluation was conducted by including lung sound noise and AWGN. The proposed model showed promising performance compared to the existing works using all evaluation criteria, even in the presence of noise. We feel that our proposed system can automatically classify CVDs based on heart auscultations, which could be used in real-world clinical scenarios in third-world countries.

## References

[1] CVD causes one-third of deaths worldwide: Study examines global burden of cvd from 1990 to 2015 - american college of cardiology.
URL https://www.who.int/en/news-room/fact-sheets/detail/cardiovascular-diseases-(cvds)
[2] A. Rosengren, A. Smyth, S. Rangarajan, C. Ramasundarahettige, S. I. Bangdiwala, K. F. AlHabib, A. Avezum, K. B. Boström, J. Chifamba, S. Gulec, et al., Socioeconomic status and risk of cardiovascular disease in 20 low-income, middle-income, and high-income countries: the prospective urban rural epidemiologic (pure) study, The Lancet Global Health 7 (6) (2019) e748–e760.
[3] S. K. Ghosh, R. Ponnalagu, R. K. Tripathy, G. Panda, R. B. Pachori, Automated heart sound activity detection from pcg signal using time–frequency-domain deep neural network, IEEE Transactions on Instrumentation and Measurement 71 (2022) 1–10.
[4] T. H. Chowdhury, K. N. Poudel, Y. Hu, Time-frequency analysis, denoising, compression, segmentation, and classification of pcg signals, IEEE Access 8 (2020) 160882–160890.
[5] A. K. Dwivedi, S. A. Imtiaz, E. Rodriguez-Villegas, Algorithms for automatic analysis and classification of heart sounds–a systematic review, IEEE Access 7 (2018) 8316–8345.
[6] R. Nersisson, M. M. Noel, Heart sound and lung sound separation algorithms: a review, Journal of medical engineering & technology 41 (1) (2017) 13–21.
[7] D. Gradolewski, G. Redlarski, Wavelet-based denoising method for real phonocardiography signal recorded by mobile devices in noisy environment, Computers in biology and medicine 52 (2014) 119–129.

[8] P. Bentley, G. Nordehn, M. Coimbra, S. Mannor, R. Getz, The pascal classifying heart sounds challenge 2011, CHSC2011 (2011).
   URL http://www.peterjbentley.com/heartchallenge/index.html
[9] G.-Y. Son, S. Kwon, Classification of heart sound signal using multiple features, Applied Sciences 8 (12) (2018) 2344.
[10] C. Liu, D. Springer, Q. Li, B. Moody, R. A. Juan, F. J. Chorro, F. Castells, J. M. Roig, I. Silva, A. E. Johnson, et al., An open access database for the evaluation of heart sound algorithms, Physiological measurement 37 (12) (2016) 2181.
[11] M. N. Homsi, P. Warrick, Ensemble methods with outliers for phonocardiogram classification, Physiological measurement 38 (8) (2017) 1631.
[12] J. Rubin, R. Abreu, A. Ganguli, S. Nelaturi, I. Matei, K. Sricharan, Classifying heart sound recordings using deep convolutional neural networks and mel-frequency cepstral coefficients, in: 2016 Computing in cardiology conference (CinC), IEEE, 2016, pp. 813–816.
[13] Y. N. Fuadah, M. A. Pramudito, K. M. Lim, An optimal approach for heart sound classification using grid search in hyperparameter optimization of machine learning, Bioengineering 10 (1) (2022) 45.
[14] H. Malik, U. Bashir, A. Ahmad, Multi-classification neural network model for detection of abnormal heartbeat audio signals, Biomedical Engineering Advances 4 (2022) 100048.
[15] F. Noman, C.-M. Ting, S.-H. Salleh, H. Ombao, Short-segment heart sound classification using an ensemble of deep convolutional neural networks, in: ICASSP 2019-2019 IEEE International Conference on Acoustics, Speech and Signal Processing (ICASSP), IEEE, 2019, pp. 1318–1322.
[16] M. Wang, B. Guo, Y. Hu, Z. Zhao, C. Liu, H. Tang, Transfer learning models for detecting six categories of phonocardiogram recordings, Journal of Cardiovascular Development and Disease 9 (3) (2022) 86.
[17] E. Kay, A. Agarwal, Dropconnected neural networks trained on time-frequency and inter-beat features for classifying heart sounds, Physiological measurement 38 (8) (2017) 1645.
[18] B. M. Whitaker, P. B. Suresha, C. Liu, G. D. Clifford, D. V. Anderson, Combining sparse coding and time-domain features for heart sound classification, Physiological measurement 38 (8) (2017) 1701.
[19] D. S. Panah, A. Hines, S. McKeever, Exploring the impact of noise and degradations on heart sound classification models, Biomedical Signal Processing and Control 85 (2023) 104932.
[20] M. Zabihi, A. B. Rad, S. Kiranyaz, M. Gabbouj, A. K. Katsaggelos, Heart sound anomaly and quality detection using ensemble of neural networks without segmentation, in: 2016 computing in cardiology conference (CinC), IEEE, 2016, pp. 613–616.
[21] T.-c. I. Yang, H. Hsieh, Classification of acoustic physiological signals based on deep learning neural networks with augmented features, in: 2016 Computing in Cardiology Conference (CinC), IEEE, 2016, pp. 569–572.
[22] C. Potes, S. Parvaneh, A. Rahman, B. Conroy, Ensemble of feature-based and deep learning-based classifiers for detection of abnormal heart sounds, in: 2016 computing in cardiology conference (CinC), IEEE, 2016, pp. 621–624.
[23] A. I. Humayun, S. Ghaffarzadegan, M. I. Ansari, Z. Feng, T. Hasan, Towards domain invariant heart sound abnormality detection using learnable filterbanks, IEEE journal of biomedical and health informatics 24 (8) (2020) 2189–2198.
[24] V. Maknickas, A. Maknickas, Recognition of normal–abnormal phonocardiographic signals using deep convolutional neural networks and mel-frequency spectral coefficients, Physiological measurement 38 (8) (2017) 1671.
[25] A. I. Humayun, S. Ghaffarzadegan, Z. Feng, T. Hasan, Learning front-end filter-bank parameters using convolutional neural networks for abnormal heart sound detection, in: 2018 40th Annual International Conference of the IEEE Engineering in Medicine and Biology Society (EMBC), IEEE, 2018, pp. 1408–1411.
[26] A. I. Humayun, M. Khan, S. Ghaffarzadegan, Z. Feng, T. Hasan, et al., An ensemble of transfer, semi-supervised and supervised learning methods for pathological heart sound classification, arXiv preprint arXiv:1806.06506 (2018).
[27] A. M. Alqudah, Towards classifying non-segmented heart sound records using instantaneous frequency based features, Journal of medical engineering & technology 43 (7) (2019) 418–430.
[28] A. M. Alqudah, H. Alquran, I. A. Qasmieh, Classification of heart sound short records using bispectrum analysis

approach images and deep learning, Network Modeling Analysis in Health Informatics and Bioinformatics 9 (2020) 1–16.

[29] S. Ghosh, P. R N, R. Tripathy, U. R. Acharya, Deep layer kernel sparse representation network for the detection of heart valve ailments from the time-frequency representation of pcg recordings, BioMed Research International (12 2020). doi:10.1155/2020/8843963.

[30] S. L. Oh, V. Jahmunah, C. P. Ooi, R.-S. Tan, E. J. Ciaccio, T. Yamakawa, M. Tanabe, M. Kobayashi, U. R. Acharya, Classification of heart sound signals using a novel deep wavenet model, Computer Methods and Programs in Biomedicine 196 (2020) 105604.

[31] N. Baghel, M. K. Dutta, R. Burget, Automatic diagnosis of multiple cardiac diseases from pcg signals using convolutional neural network, Computer Methods and Programs in Biomedicine 197 (2020) 105750.

[32] W. Zeng, Z. Lin, C. Yuan, Q. Wang, F. Liu, Y. Wang, Detection of heart valve disorders from pcg signals using tqwt, fa-mvemd, shannon energy envelope and deterministic learning, Artificial Intelligence Review (2021) 1–38.

[33] S. B. Shuvo, S. N. Ali, S. I. Swapnil, M. S. Al-Rakhami, A. Gumaei, Cardioxnet: A novel lightweight deep learning framework for cardiovascular disease classification using heart sound recordings, IEEE Access 9 (2021) 36955–36967.

[34] M. Alkhodari, L. Fraiwan, Convolutional and recurrent neural networks for the detection of valvular heart diseases in phonocardiogram recordings, Computer Methods and Programs in Biomedicine 200 (2021) 105940.

[35] S. Tiwari, A. Jain, A. K. Sharma, K. M. Almustafa, Phonocardiogram signal based multi-class cardiac diagnostic decision support system, IEEE Access 9 (2021) 110710–110722.

[36] M. A. Kobat, S. Dogan, Novel three kernelled binary pattern feature extractor based automated pcg sound classification method, Applied Acoustics 179 (2021) 108040.

[37] Y. Al-Issa, A. M. Alqudah, A lightweight hybrid deep learning system for cardiac valvular disease classification, Scientific Reports 12 (1) (2022) 14297.

[38] Icbhi 2017 challenge, accessed: 2023-02-11 (2017).
URL https://bhichallenge.med.auth.gr/ICBHI_2017_Challenge

[39] S. E. Schmidt, C. Holst-Hansen, C. Graff, E. Toft, J. J. Struijk, Segmentation of heart sound recordings by a duration-dependent hidden markov model, Physiological measurement 31 (4) (2010) 513.

[40] A. Raza, A. Mehmood, S. Ullah, M. Ahmad, G. S. Choi, B.-W. On, Heartbeat sound signal classification using deep learning, Sensors 19 (21) (2019) 4819.

[41] S. Reichert, R. Gass, C. Brandt, E. Andrès, Analysis of respiratory sounds: state of the art, Clinical medicine. Circulatory, respiratory and pulmonary medicine 2 (2008) CCRPM–S530.

[42] A. Meintjes, A. Lowe, M. Legget, Fundamental heart sound classification using the continuous wavelet transform and convolutional neural networks, in: 2018 40th annual international conference of the IEEE engineering in medicine and biology society (EMBC), IEEE, 2018, pp. 409–412.

[43] S. Debbal, F. Bereksi-Reguig, Analysis of the second heart sound using continuous wavelet transform, Journal of medical engineering & technology 28 (4) (2004) 151–156.

[44] N. Gautam, S. B. Pokle, Wavelet scalogram analysis of phonopulmonographic signals, International Journal of Medical Engineering and Informatics 5 (3) (2013) 245–252.

[45] S. B. Shuvo, S. N. Ali, S. I. Swapnil, T. Hasan, M. I. H. Bhuiyan, A lightweight cnn model for detecting respiratory diseases from lung auscultation sounds using emd-cwt-based hybrid scalogram, IEEE Journal of Biomedical and Health Informatics 25 (7) (2020) 2595–2603.

[46] S. Davis, P. Mermelstein, Comparison of parametric representations for monosyllabic word recognition in continuously spoken sentences, IEEE transactions on acoustics, speech, and signal processing 28 (4) (1980) 357–366.

[47] K.-W. Ha, J.-W. Jeong, Motor imagery eeg classification using capsule networks, Sensors 19 (13) (2019) 2854.

[48] S. Krishnan, Biomedical signal analysis for connected healthcare, Elsevier, 2021.

[49] C. Schörkhuber, A. Klapuri, Constant-q transform toolbox for music processing, in: 7th sound and music computing conference, Barcelona, Spain, 2010, pp. 3–64.

[50] B. Xu, N. Wang, T. Chen, M. Li, Empirical evaluation of rectified activations in convolutional network, arXiv preprint arXiv:1505.00853 (2015).

[51] F. N. Iandola, S. Han, M. W. Moskewicz, K. Ashraf, W. J. Dally, K. Keutzer, Squeezenet: Alexnet-level accuracy with 50x fewer parameters and¡ 0.5 mb model size, arXiv preprint arXiv:1602.07360 (2016).

[52] C. Ding, Y. Jia, G. Cui, C. Chen, X. Zhong, Y. Guo, Continuous human activity recognition through parallelism lstm with multi-frequency spectrograms, Remote Sensing 13 (21) (2021) 4264.

[53] N. Kannathal, U. R. Acharya, C. M. Lim, P. Sadasivan, S. Krishnan, Classification of cardiac patient states using artificial neural networks, Experimental & Clinical Cardiology 8 (4) (2003) 206.

[54] F. B. Azam, M. I. Ansari, S. I. S. K. Nuhash, I. McLane, T. Hasan, Cardiac anomaly detection considering an additive noise and convolutional distortion model of heart sound recordings, Artificial Intelligence in Medicine 133 (2022) 102417.

[55] S. S. Alam, A. Chakma, M. H. Rahman, R. Bin Mofidul, M. M. Alam, I. B. K. Y. Utama, Y. M. Jang, Rf-enabled deep-learning-assisted drone detection and identification: An end-to-end approach, Sensors 23 (9) (2023). doi:10.3390/s23094202.
URL https://www.mdpi.com/1424-8220/23/9/4202

[56] J. Acharya, A. Basu, Deep neural network for respiratory sound classification in wearable devices enabled by patient specific model tuning, IEEE transactions on biomedical circuits and systems 14 (3) (2020) 535–544.

[57] S. Y. Nikouei, Y. Chen, S. Song, R. Xu, B.-Y. Choi, T. R. Faughnan, Real-time human detection as an edge service enabled by a lightweight cnn, in: 2018 IEEE International Conference on Edge Computing (EDGE), IEEE, 2018, pp. 125–129.

[58] H. W. Loh, C. P. Ooi, S. Seoni, P. D. Barua, F. Molinari, U. R. Acharya, Application of explainable artificial intelligence for healthcare: A systematic review of the last decade (2011–2022), Computer Methods and Programs in Biomedicine 226 (2022) 107161. doi:https://doi.org/10.1016/j.cmpb.2022.107161.
URL https://www.sciencedirect.com/science/article/pii/S0169260722005429